# Finding the white male: The prevalence and consequences of algorithmic gender and race bias in political Google searches

Tobias Rohrbach [1], Mykola Makhortykh [1] und Maryna Sydorova [1]

**Abstract:** Search engines like Google have become major information gatekeepers that use artificial intelligence (AI) to determine who and what voters find when searching for political information. This article proposes and tests a framework of algorithmic representation of minoritized groups in a series of four studies. First, two algorithm audits of political image searches delineate how search engines reflect and uphold structural inequalities by under- and misrepresenting women and non-white politicians. Second, two online experiments show that these biases in algorithmic representation in turn distort perceptions of the political reality and actively reinforce a white and masculinized view of politics. Together, the results have substantive implications for the scientific understanding of how AI technology amplifies biases in political perceptions and decision-making. The article contributes to ongoing public debates and cross-disciplinary research on algorithmic fairness and injustice.

**Keywords:** Descriptive representation, Gender/race bias, Algorithmic audit, Strategic discrimination, Artificial intelligence

Last update: *25 April 2024*

**This pre-print is currently being peer-reviewed.**

Contact authors for supplemental materials.

1 University of Bern, Institute of Communication and Media Research, Fabrikstrasse 8, 3012 Bern, Switzerland, tobias.rohrbach@unibe.ch, https://orcid.org/0000-0002-0151-1377; mykola.makhortyk@unibe.ch, https://orcid.org/0000-0001-7143-5317; maryna.sydorova@unibe.ch, https://orcid.org/0000-0000-0000-0000

Though political landscapes around the world are slowly changing, the face of contemporary political decision-making is still disproportionately male and, in the Global North, disproportionately white. In 2024, global descriptive representation for women in parliaments is 26.7% (Inter-parliamentary Union 2024). In the U.S. House of Representatives, 29% of members are women and 26% are non-white,[1] making the 118th Congress the most ethnically diverse and gender-balanced to date in US history (Pew Research Center 2023a,b). Yet according to census data, the shares of women and non-white persons in the general US population are 50.5% and 41.1% respectively (U.S. Census Bureau 2023). Decades of scientific exploration into the mechanisms of the underrepresentation of minoritized groups indicate that political structures, gender perceptions, and the media environment are multicausal drivers of gender and race inequalities in politics (Bratton and Ray 2002; Bos et al. 2022; Dolan 1998; Kahn 1994; Kanthak and Woon 2015; Mendelberg et al. 2014; Thomsen and King 2020; Teele et al. 2018; Dolan and Hansen 2018; Huddy and Terkildsen 1993; O'Brien 2015). In this article, we turn to artificial intelligence (AI) driven search engines like Google as a novel and so far underexplored driver of political exclusion of underrepresented groups. Specifically, we propose and test a framework of algorithmic representation to delineate the role of search engines in constructing masculinized and white views of politics as well as delineate the consequences of such views for political perceptions.

Search engines have become major information gatekeepers that use artificial intelligence (AI) to determine who and what voters find when searching for political information (Trielli and Diakopoulos 2022; Urman and Makhortykh 2022; Wallace 2018; White and Horvitz 2015). How information is represented on search engines has been shown to influence perceptions of political campaigns (Epstein and Robertson 2015; Zweig 2017) and individual vote choices (Diakopoulos et al. 2018). Selection and ranking of information by search engines—and, crucially, biases of such algorithmic curation,[2] including the systematic under- and misrepresentation of gender and racial groups (Makhortykh et al. 2021; Noble 2018; Urman and Makhortykh 2022)—in turn shape perceptions of political realities. Specifically, search engine outputs that underrepresent women or non-white politicians may reinforce

---

[1] We use the term "non-white" in equivalence to the U.S. Census Bureau's category of "White alone (not Latino or Hispanic)" to refer to individuals who do not report as being of exclusively white Caucasian origin.

[2] We understand bias in relation to computer systems as system performance that is systematically skewed towards particular individuals or group (see Friedman and Nissenbaum 1996)

inequalities in politics by reifying the collective stereotypical representation of politicians as white and male (Bateson 2020; Corbett et al. 2022; Stokes-Brown and Dolan 2010; Vlasceanu and Amodio 2022; Philpot and Walton 2007).

Theoretically, our framework of *algorithmic representation* integrates established literature on descriptive representation and gendered mediation with more recent work on strategic discrimination to showcase how algorithmic biases systematically shift perceptions of political realities in ways that exacerbate existing gender and race inequalities. We test causal relationships by combining evidence from two observational algorithm auditing studies with two survey experiments.

We provide compelling evidence supporting our assumptions regarding algorithmic representation in the context of politics. In our algorithmic audits (studies 1 and 2), we find that women are consistently underrepresented in Google searches dealing with political information in 56 countries and for both lower and upper legislative bodies. Moreover, this algorithmic underrepresentation moderately correlates with women's actual descriptive representation in countries' legislative bodies. The experimental analysis (studies 3 and 4) shows that underrepresentation of women or non-white politicians in Google search outputs leads voters to underestimate these groups' descriptive representation by roughly 10 percentage points. Crucially, mediation analyses suggest that this perceptual bias regarding descriptive representation results in undesirable political perceptions concerning the viability of politicians from minoritized groups.

The results advance scientific understanding of how AI gatekeepers reflect structural inequalities in politics—and how they amplify them by exacerbating biases in political perceptions and decision-making. Our framework of algorithmic representation adds a crucial component to understanding the enduring structural disadvantages of women and people of color in an increasingly AI-driven political landscape. It contributes to current public and interdisciplinary scientific research concerning algorithmic fairness and injustice (Birhane 2021; Kalluri 2020; Wong 2020; Weinberg 2022). Such insight is integral for raising societal awareness of the discriminatory tendencies inherent in AI-driven systems within increasingly digital political spaces (Friesen et al. 2021). Moreover, It provides an empirical basis for developing new regulation for preventing risks associated with the growing adoption of AI and is thus relevant for a broad range of stakeholders, including policymakers and industry representatives, but also civil society and human right advocacy groups.

## SEARCH ENGINES AND WOMEN'S DESCRIPTIVE REPRESENTATION

Collective descriptive representation is the quality of political institutions to represent their citizens on the basis of shared political and sociodemographic characteristics (Atkeson and Carrillo 2007; Bratton and Ray 2002). Increases in descriptive representation of minoritized groups have been connected to policy outcomes that benefit their substantive interests, thus decreasing structural inequalities (Bratton and Ray 2002; Hessami and da Fonseca 2020; Lowande et al. 2019; Wängnerud 2009; Mendelberg et al. 2014). The inclusion of women in governing bodies also has a symbolic quality (Stokes-Brown and Dolan 2010). Whether governments adequately approximate the demographic composition of their population "sends important signals about who should (and should not) participate in politics and the degree to which certain groups and interests will receive a fair hearing" (Stauffer 2021, 1226). Descriptive representation gives rise to a range of political perceptions, such as impressions of who is qualified to be a political leader and more symbolic ways of feeling (un)fairly represented (Atkeson and Carrillo 2007). At its very core, descriptive representation therefore acts as a visibility mechanism that highlights political perspectives with which citizens can identify (Dolan 1998).

Scholars have identified media discourse as an important factor shaping how citizens see themselves and whether they perceive their interests as represented (Kahn 1994). The prevailing conclusion of this line of work is that political coverage under- and misrepresents minoritized groups (Gershon 2013; Van Der Pas and Aaldering 2020) and that these distortions in coverage result in diminished evaluations of political candidates belonging to these groups (Rohrbach et al. 2023; van Oosten et al. 2024). In this article, we tackle this much needed update of this line of reasoning and propose a framework of algorithmic representation. The core argument of the framework is that political inequalities are built into search algorithms whose output then reinforces political perceptions that cement disadvantages in the political landscape (for a similar argument see Vlasceanu and Amodio 2022). In the following, we formulate our framework as a set of four types of bias driving the extent of algorithmic representation and its consequences for political perceptions.

**Algorithmic representation as outcome of structural inequalities**

Most people never encounter politicians in person but through intermediaries, such as media coverage (Shehata and Strömbäck 2014; Kahn 1994). The gendered mediation thesis posits that the transformation of a political reality into a mediated reality happens through various journalistic filters (Gidengil and Everitt 2000; Trimble et al. 2022). These filters take the form of gendered organizational structures of newsrooms, journalists' sourcing and writing practices and their professional norms (Riedl et al. 2022). This filtered mediation presents audiences with coverage that quantitatively and qualitatively upholds a masculinized (and white) view of politics. These views matter as they mark the core of gendered political socialization (Bos et al. 2022; Dassonneville and McAllister 2018) and ultimately shape women's political ambition to emerge as candidates (Fox and Lawless 2011; Kanthak and Woon 2015).

Today, indirect encounters with political figures increasingly take place in AI-curated digital spaces. AI powered search engines—but also social media feeds and news recommender systems—structure what accounts of politics voters get to see (and which they do not; see e.g. Wallace 2018; Friesen et al. 2021). In "post-broadcast democracies", the human gatekeeping influence of journalists is increasingly substituted for AI gatekeeping (see Stier et al. 2022). Studies show that search engines like Google have become key sources of political information (Trevisan et al. 2018; Urman and Makhortykh 2023). Moreover, in some cases, people place more trust in the information obtained through their internet searches than in traditional news media (see, e.g., Edelman Trust Institute 2024).

The widespread use and perceived trustworthiness of search engines are cause for concern, as research has documented that such algorithmic systems can be biased in political contexts (Epstein and Robertson 2015; Trielli and Diakopoulos 2022). A number of studies have specifically documented the "representational harm" (Fabris et al. 2020, 5) arising from search engines' tendency to underrepresent women (Pradel 2021; Vlasceanu and Amodio 2022) as well as people of color (Noble 2018; Makhortykh et al. 2021). Applying the filtering logic of the gendered mediation thesis to search engines, we thus expect that political searches on Google have a built-in bias against women (*H1*):

> *H1: Baseline bias:* Women are algorithmically underrepresented in (gender-neutral)

political Google image search outputs.[3]

If the masculinized political and social reality translates into an algorithmic underrepresentation of women on Google, then the extent of this underrepresentation should vary across societies with different political benchmarks regarding women's descriptive representation.[4] Borrowing from social role theory, Niven (1998, 2004) explains women's under-/misrepresentation in political media coverage as a function of the paucity of women leadership roles. This view is in line with dynamic stereotype models which emphasize that stereotypes are shaped by distributions of women and men among distinct occupational roles—that is, in politics, their presence in office (Diekman et al. 2004; Eagly et al. 2020; van der Pas et al. 2023). Not unlike humans, search engine algorithms should be able to observe the distribution of women and men into occupational roles by more frequently encountering and indexing content showing women politicians in countries with higher descriptive representation. Yet due to the inherent baseline bias in their performance (see *H1*), we expect search engine algorithms to imperfectly mirror gendered role distributions by underrepresenting women relative to their descriptive representation (*H2*).

> *H2: Distribution bias:* Women's algorithmic underrepresentation is positively correlated with their descriptive representation across countries but does not exceed it.

## Algorithmic representation as a driver of political perceptions

So far we have established the hypotheses that women are likely to be algorithmically rendered less visible than men in the output of political Google searches across national contexts. Drawing on strategic discrimination literature, we now outline the consequences of algorithmic representation as two causally linked biases influencing political perceptions. Bateson's (2020) theory of strategic discrimination describes the tendency of voters to withhold electoral support for minority candidates not on the basis of their own direct prejudice but out of a strategic calculus of these candidates' perceived ability to win the election (see also Ashworth et al. 2024). While direct race or gender bias in the electorate may be declining (Juenke and Shah 2016; Schwarz and Coppock 2022; Teele et al. 2018), studies on strategic discrimination have consistently found that participants perceive women and Black candidates as less electable and, consequently, are less likely to vote for them (Bateson 2020; Corbett et al. 2022; Green et al. 2022).

---

[3] For the first two hypotheses, we focus exclusively on gender and disregard the question and disregard race for two reasons. First, measuring the share of non-white politicians in Google output would require labeling the race of depicted persons on a purely visual basis. In addition to strong ethical concerns, computational labeling approaches rely on face recognition algorithms which are themselves subject to inherent race biases (Cavazos et al. 2021; Scheuerman et al. 2019). Second, governments do not typically provide information about the racial or ethnic background of their legislative branches, making it impossible to establish a benchmark for comparison.

[4] Note that this argument differentiates between countries with a history of strong inclusion of women and countries with policy-induced gender parity due to gender quotas.

We theorize that algorithmic representation operates within the mechanism of strategic discrimination in two steps. First, the representation of minoritized groups in search engine outputs distorts individual perceptions of descriptive representation of those groups (*H3*). We thus posit that search engine output serves as a political heuristic feeding the gendered perception gap (Lau and Redlawsk 2001; McDermott 1997), which is the "systematic overestimation by men and underestimation by women of their electability" (Ashworth et al. 2024, 290).As most voters have no clear idea of the actual demographic composition of political institutions (see, e.g., Stauffer 2021), search engine users will approximate the algorithmically curated reality with the political reality. Such misperceptions matter, as "public opinion on descriptive representation is likely to explain political behavior at both the individual and systemic levels" (Dolan and Sanbonmatsu 2009, 410; Dolan and Hansen 2018). This suggests a circular logic in which political inequalities are coded into search engines whose biases then reinforce the skewed perceptions of the political status quo.

> *H3: Perceptual bias:* Algorithmic underrepresentation of women and non-white politicians lowers estimations of their descriptive representation.

Second, these algorithmically driven misperceptions regarding the actual inclusion of minoritized groups affect electability assessments (*H4a*). Bateson (2020) identifies the perceived ability of a candidate to raise funds, to generate media coverage and to win over voters as determinants of electability. Voter perceptions of these dimensions remain vague and difficult to empirically capture. We therefore broaden this approach and introduce the representation in the output of Google searches as a key variable in voters' strategic calculus. Because search engines are so regularly and widely used as well as trusted, they present a particularly plausible source for individuals to form their "second-order preferences"—that is, the perception of the ideological and demographic candidate profile the abstract median voter would electorally support (Green et al. 2022, 888). Furthermore, information searches online are closer to individuals' daily life than, for instance, concerns about politicians' fundraising capabilities.

In addition to lowering perceived electability of candidates belonging to minoritized groups, there is reason to believe that underestimations of descriptive representation also diminishes voters' external efficacy, understood as the perception that government is responsive to their concerns (*H4b*; Atkeson and Carrillo 2007). Specifically, Stauffer (2021) convincingly showed for the case of women politicians that it is voters' subjective perception of women's inclusion rather than their objective reality that shapes political evaluations (see also Dolan and Sanbonmatsu 2009). We replicate this argument and extend it to non-white politicians.

> *H4: strategic bias:* Underestimated descriptive representations in turn lower voters' evaluations of (a) the electability of women and non-white politicians and (b) voters' external efficacy.

## MEASURING THE EXTENT OF ALGORITHMIC REPRESENTATION

### Study 1: Auditing gender bias in Google image search

To investigate women's algorithmic representation (*H1, H2*), we conducted an algorithm audit of Google image searches. Algorithm auditing has become the dominant research method for "diagnosing problematic behavior in algorithmic systems" (Bandy 2021, 1). Such audits often involve the creation of virtual agents used to simulate the user interaction with an algorithmic system under controlled conditions and to generate system outputs (Urman et al. 2022, 2024).

The audit focuses on Google image searches for several reasons. First, analyses of Google trends have linked voters' information search behavior to political events and election outcomes, thus highlighting its relevance of the search engine for political decision-making (Stephens-Davidowitz 2014; Trevisan et al. 2018; Urman and Makhortykh 2023). Second, longitudinal analyses indicate that the average citizen is spending less time reading and more time viewing images (American Academy of Arts & Sciences 2019). Politicians in turn respond to voters' visual preference by increasingly relying on visual communication for their political image building (Bast 2024; Carpinella and Bauer 2021). Third, a recent large-scale study on representation of professional occupations in Google text and image searches suggests that images were particularly likely to prime and amplify gender bias and thus "come at a critical social cost" (Guilbeault et al. 2024, 6).

For the first study, we manually simulated user activity in 56 different countries with a local IP address (accessed through a VPN service called Le VPN; `https://www.le-vpn.com/`) in the first week of August 2023.[5] The simulation consisted of deploying a virtual agent—that is, a computer script—that was automated to conduct Google image search queries in in each geographical location. For consistency, we used the ".com" version of Google due to the possibility of not having a language-specific version of the search engine for all cases. We conducted a Google image search query for the country's lower or single chamber of parliament with the following pattern: [name of legislative body][person] (see Table B1 in the Supplemental Materials). For bicameral systems, the query was repeated for the upper chamber as well. In line with Vlasceanu and Amodio (2022), we used the country's dominant language to conduct the query and added the term "person" to obtain results related to people rather than buildings. Moreover, the abstract term "person" carries no or only very weak gendered connotations in most languages whereas more specific terms such as citizen, politician, or member of parliament would require choosing a grammatical gender (for a discussion see Vlasceanu and Amodio 2022). For the United States, the queries thus read: "house of representatives person" and "senate person". For each conducted query, we collected the first 75 images and extracted the number of persons and their gender by means of computer vision using the commercial Amazon Rekognition platform (`https://aws.amazon.com/rekognition/`).[6] This yielded a data set of 6,363 images

---

[5] All studies were pre-registered and received ethical clearance from the institutional review board. See section A in the Supplemental Materials for more information.

[6] Amazon Rekognition predicts the gender of a depicted person as a male vs. female binary based on physical appearance. An algorithmic system itself, Amazon Rekognition's gender prediction has been shown to work best for (white) cis-gender women and men with true positive rates of 95% and 99% respectively (Scheuerman

depicting 58,343 persons. We then merged the data at the country-level with women's actual descriptive representation in these legislative bodies (Inter-parliamentary Union 2024). We used the clean browser and cleared its history after each query to prevent the possible impact of the browser history on search outputs.

Our main measure of algorithmic representation is the share of women of all depicted persons in each image (e.g., 50% in an image with two women and two men). We additionally repeat all analyses with the absolute number of depicted women and men as well as a dummy variable predicting the presence of at least one man or woman in each image.

We first test our expectations that search engine algorithms have a baseline bias that results in women's *absolute* underrepresentation in Google image outputs compared to men (*H1*). We run generalized linear mixed effects models with by-country random intercepts to predict the share, presence and number of women in output images. In line with our expectation, the results show a consistent algorithmic underrepresentation of women on all three measures and in search queries for both lower and upper chambers (see Table 1). Of note, the extent of women's algorithmic underrepresentation (lower: 29.2%, *CI*[26.8%-31.7%]; upper: 29.1%, *CI*[26.8%-31.5%]) lies just a few percentage points above their average global descriptive representation of 26.7%.

Next we turn to our hypothesis that women's algorithmic representation tracks with the gendered distribution into the political roles, as measured by women's descriptive representation (*H2*). For this, we assess the correlation between women's algorithmic and actual descriptive representation. We find significant positive correlations in both chambers (lower: $r(56) = 0.37$, $p = 0.005$; upper: $r(28) = 0.52$, $p = 0.004$). These associations indicate that the proportion of women in Google images is higher (lower) in countries and chambers with more (fewer) elected women (see Figure 1A).[7]

Contrary to expectation, the evidence does not indicate a clear pattern of *relative* algorithmic repre-

---

et al. 2019). Commercial gender detection models notoriously perform worse for women and non-white people but retain accuracy rates of 80% and better (Albiero et al. 2020; Schwemmer et al. 2020). We compared the automatically annotated gender with a (binary) manual classification conducted by the authors ($n = 300$) and achieved satisfying reliability (Krippendorff's $\alpha = 0.92$).

[7]These bivariate associations hold even after controlling for country- or query-level predictors (see section C1 in the Supplemental Materials for robustness checks of this finding).

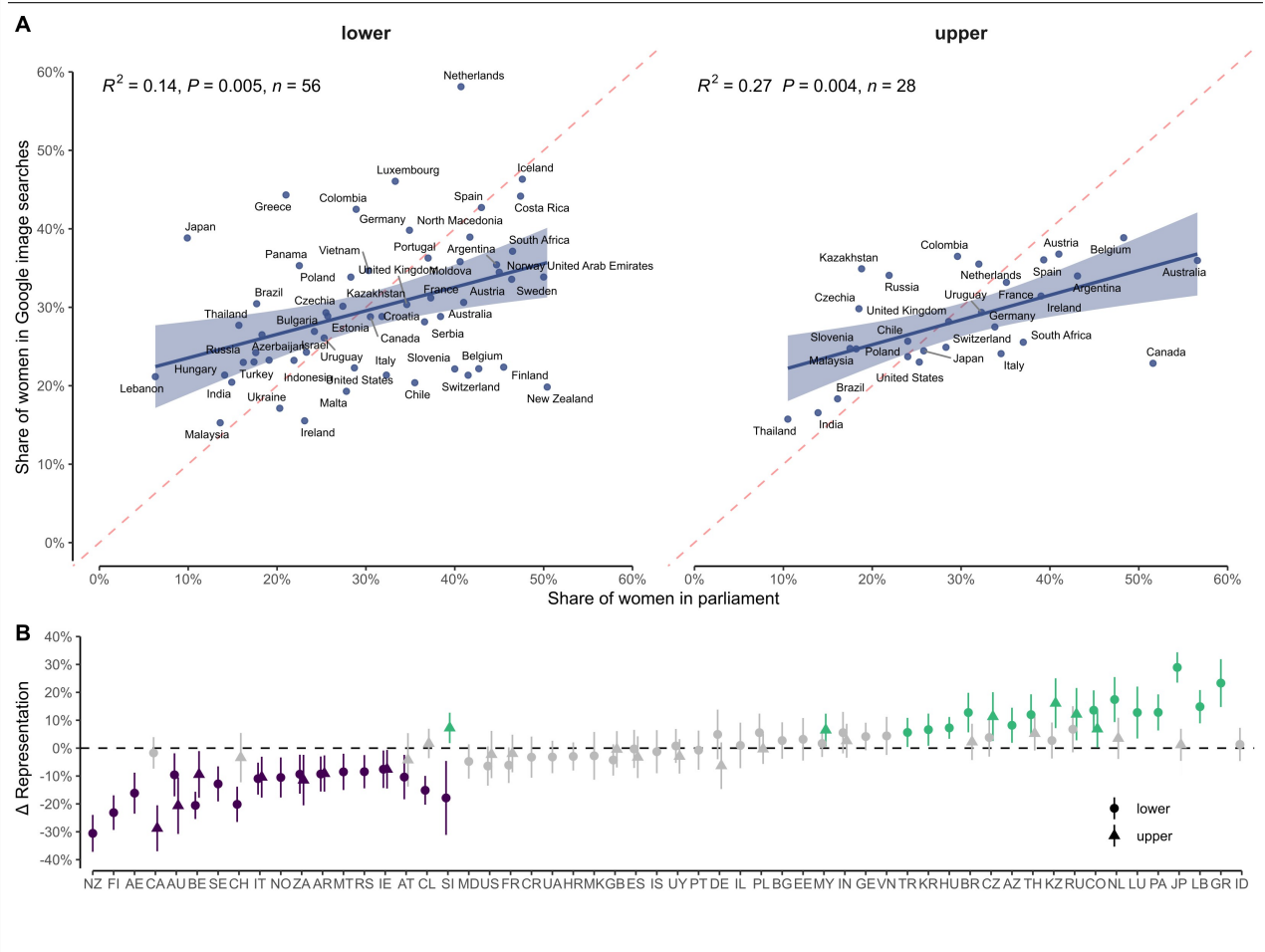

**FIGURE 1. Associations between women's algorithmic and descriptive representation in parliamentary chambers across 56 countries**

*Note:* Panel A depicts women's algorithmic (y-axis) and descriptive representation (x-axis; Inter-parliamentary Union, 2024) in lower and upper parliamentary chambers across 56 countries in study 1. Panel B depicts the mean differences in representation along with 95% confidence intervals (y-axis) per country and chamber. Positive (negative) values indicate that women's algorithmic representation is higher (lower) than their descriptive representation. See Table C1.2 in the Supplemental Materials for full results.

sentation. Figure 1B illustrates the differences between women's actual and algorithmic representation for all queries. We find no significant difference for the majority of countries and chambers (e.g., the U.S. House of Representatives), indicating that search engines accurately mirror women's inclusion in most legislative institutions. Google's search algorithm introduces bias in women's underrepresentation in 20 cases. For instance, there is a relative underrepresentation of women in the output for the U.S. Senate query by -5.8% percentage points (*CI*[-11.6%- -0.07%]). However, women are algorithmically overrepresented relative to their descriptive representation in 21 cases.

## Study 2: Internal replication and robustness check

Study 2 constitutes an internal replication of the first study with the goal to probe the robustness of the previous audit (a) for the time of data collection and (b) the noise due to randomization of search engine outputs (Haim et al. 2017; Urman et al. 2022). Data collection for study 2 included queries for 20 legislative bodies (nine bicameral and two unicameral countries) and took place in March 2024, almost a year after the first study. We deployed 20 virtual agents—rather than just one—which simultaneously conducted the same queries to account for random noise in each Google search. To account for geographical influences on Google searches, we paralleled this procedure in each of the eleven countries by modeling the location of the virtual agents through the set of IP addresses provided by Google Compute Engine.[8] For each query we collected and coded up to the first 50 images in the Google search output. This resulted in a data set of 152,098 images depicting 1,324,560 persons.

For the analysis we repeat the generalized linear mixed effects models from the previous study with random intercepts for the country of the query, the location in which the agent performed the query, and the agent id. We first test the baseline bias in women's representation on Google image searches (*H1*). Confirming results from the previous study, women are underrepresented on all measures (see Table 1). For example, women account for 22.5% (*CI*[18.2%-26.9%]) of depicted persons in searches for lower parliamentary chambers and 25.0% (*CI*[20.2%-29.8%]) in those for upper chambers. Next we assess whether the distribution of women's representation on Google image search outputs mirrors their actual inclusion in governments across countries (*H2*). We find that a percentage point increase in women's descriptive representation results in an increase in their algorithmic representation by half a percentage point *(b = 0.50, CI*[0.46:0.53], *p* < 0.001; see Table C2.2 in the Supplemental Materials). Mirroring the bivariate analysis from study 1, we find a positive correlation between women's algorithmic and descriptive representation across 20 legislative bodies (*r*(20) = 0.33, *p* = 0.14), though this bivariate association is not statistically significant.

---

[8]We find substantial random variation in search engine behavior. Yet the pattern of women's consistent algorithmic underrepresentation holds across VPN locations and individual agent queries. We report more detailed analyses in section C2 of the Supplemental Materials.

**TABLE 1. Panel Overview of women's algorithmic representation on Google image search outputs**

|  | Study 1 | | Study 2 | |
| --- | --- | --- | --- | --- |
|  | Lower/ | Upper | Lower | Upper |
| Share women | 29.2 | 29.1 | 22.5 | 25.0 |
|  | (26.8-31.7) | (26.8-31.5) | (18.2-26.9) | (20.2-29.8) |
| Presence women | 44.5 | 46.1 | 41.2 | 39.2 |
|  | (40.8-48.2) | (41.2-50.7) | (36.1-46.4) | (33.0-45.4) |
| Presence men | 61.4 | 67.0 | 70.3 | 67.9 |
|  | (57.7-65.2) | (63.6-70.4) | (64.1-76.5) | (59.4-76.3) |
| Number women | 2.9 | 2.4 | 3.1 | 1.3 |
|  | (2.4-3.4) | (1.9-2.8) | (2.2-4.0) | (0.9-1.7) |
| Number men | 6.9 | 5.4 | 8.1 | 3.6 |
|  | (5.6-8.2) | (4.3-6.6) | (5.6-10.6) | (2.6-4.5) |
| $N_{images}$ / $k_{countries}$ | 4243 / 56 | 2120 / 28 | 92213 / 11 | 59885 / 9 |

*Note:* Numbers are average predictions for the fixed effect intercepts from separate models per representation measure. All models are clustered around the country level; models in study 2 are additionally clustered around the location of the VPN ($k_{location}$ = 11) and the virtual agents ($k_{agent}$ = 220). See Table C1.1 in the Supplemental Materials for full results.

## EXPERIMENTAL EFFECTS OF ALGORITHMIC REPRESENTATION

### Study 3: Consequences of algorithmic gender bias

So far, our observational studies have established that women are algorithmically underrepresented in Google image searches and that this underrepresentation tends to mirror the descriptive (under)representation trends. To investigate the consequences of exposure to algorithmic underrepresentation, we conducted an online experiment where we varied the gender composition of Google search output to be either balanced (1:1) or biased against women (1:9).

We recruited a US ($n$ = 304) and a UK ($n$ = 303) sample through Prolific. Both samples are gender balanced (see section B2 in the Supplemental Materials for a breakdown of sample characteristics as well as a priori sample size calculations).

We constructed the stimuli to resemble the first 11 images of actual Google searches as closely as possible (for a similar approach see Vlasceanu and Amodio 2022). Both stimuli show screenshots of a Google image search output, keeping the same layout and search bar. The stimuli only differed in the gender composition of depicted persons. For the manipulation of depicted persons, we used nine images of persons that we collected for the Swiss, and Australian queries in study 1. We matched the images to depict either a query for the House of Representatives (US sample) or the House of Commons (UK sample). Image selection for each treatment prioritized similarities in content and context, aside from gender composition. To bolster external validity, we incorporated an image of the vacant House of Representatives (or House of Commons) and another of the institution's official flag. To prevent participant skipping, a 20-second mandatory exposure was enforced (the used stimuli are included in section B3 in the Supplemental Materials).

We measured three outcomes. First, we ask participants to provide a pre- and post-treatment *estimate of descriptive representation* of women politicians using a slider scale from 0 to 100. Second, participants rated the *perceived electability* of women compared to that of men using a single slider scale ranging from -5 (women less electable) to +5 (women more electable to win). Finally, we assessed participants' *external efficacy* beliefs by asking their agreement to two statements on a 5-point scale: (1) "Government officials care what people like me think" and (2) "People like me have a say in what the government does". We include the following measures as statistical controls: *participant gender*, *age*, *search engine use* ("If you think about a regular day, how many times do you use search engines (like Google or Bing) on a given day?"), *ideology* (two items from 1 left/liberal to 10 right/conservative), and *party vote* ("If there were elections today, which party would you vote for?").

We first assess our perceptual bias hypothesis (*H3*) that exposure to algorithmic underrepresentation distorts perceptions of political realities. In both samples, we predict participants' estimations of women's descriptive representation in a generalized linear mixed effects model with output condition (gender-balanced vs. underrepresentation) and time (pre- vs. post-treatment) as fixed effects and with by-participant random intercepts. Our hypothesis was supported by a significant condition by time interaction in both samples (U.S.: $b = 11.06$, *SE* = 0.72, $t = 15.34$, $p < 0.001$; UK: $b = 11.67$, *SE* = 0.57, $t = 20.37$, $p < 0.001$). This interaction is illustrated in Figure 2A, showing that participants have reasonable baseline perceptions of women's inclusion in politics prior to treatment. Whereas exposure to gender-balanced search engine output led to slight overestimations of women's actual representation, the algorithmic underrepresentation condition resulted in substantial underestimations of their inclusion in politics. Figure 2B shows the distribution of the net perceptual bias, calculated by subtracting women's actual descriptive representation from participants' posttreatment estimations. Participants in

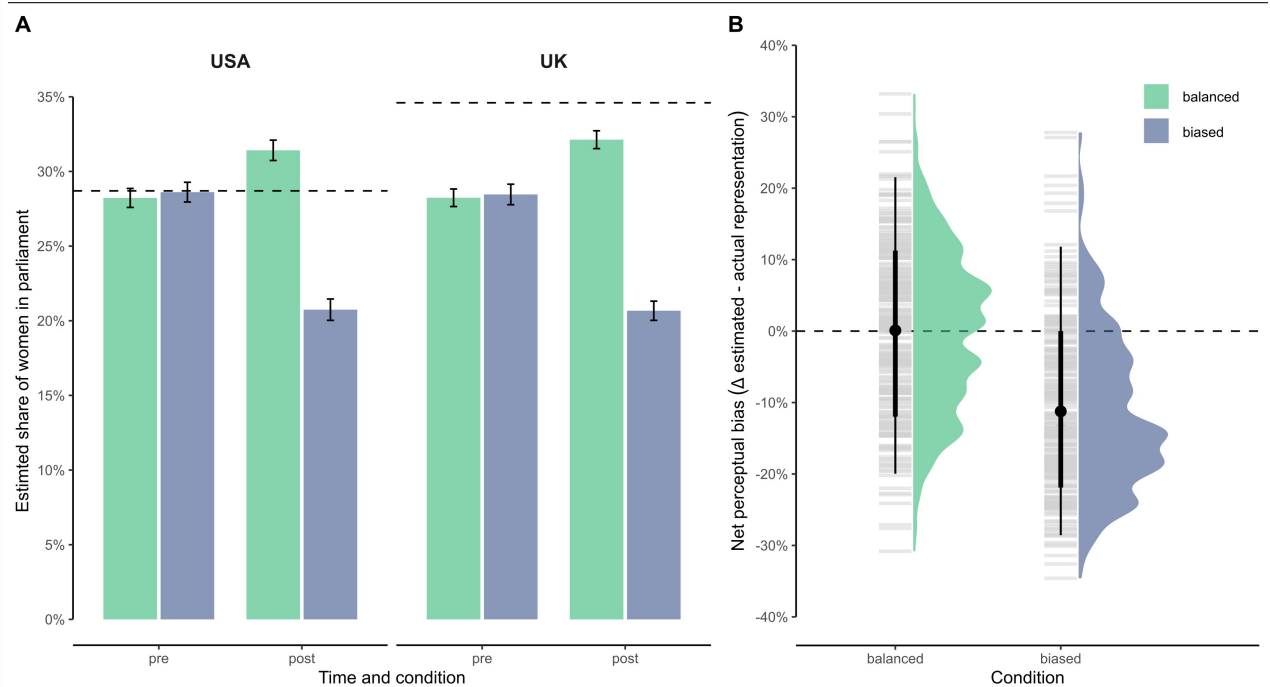

**FIGURE 2.** Experimental effects of exposure to algorithmic representation on women's estimated descriptive representation in study 3

*Note:* Panel A shows the predicted pre-and posttreatment estimations of women's descriptive representation in both experimental conditions and samples. The dashed lines denote women's actual descriptive representation. Panel B depicts the distribution of the net perceptual bias, calculated by subtracting women's actual descriptive representation from participants' posttreatment estimations by experimental condition. See Table C3.1 in the Supplemental Materials for full results.

the algorithmic underrepresentation on average misjudged women's actual representation in the lower chamber by 10.9 percentage points ($SD = 12.1$), compared to the equal representation condition ($M = 0.09$, $SD = 11.3$), $t(588) = 11.4$, $p < 0.001$, Hedges' $g = 0.93$.

Next, we test the strategic bias hypothesis by conducting simple mediation analyses (see Figure 3). For that purpose, we estimate the indirect effect of algorithmic underrepresentation (vs. balanced) on participants' electability assessments of women politicians (*H4a*) as well as their own external efficacy (*H4b*) through the (mis)estimation of women's descriptive representation. We find significant negative average causal mediation effects (*ACME*) for both outcomes (electability: -0.17, *CI*[-0.34:-0.03]; external efficacy: -0.19, *CI*[-0.28:-0.11]). Together, the evidence shows that exposure to biases in search engine output drastically lowers voters' baseline perceptions of the role of women in political realities which in turn present a key source of womens' electability assessments and voters' efficacy beliefs.

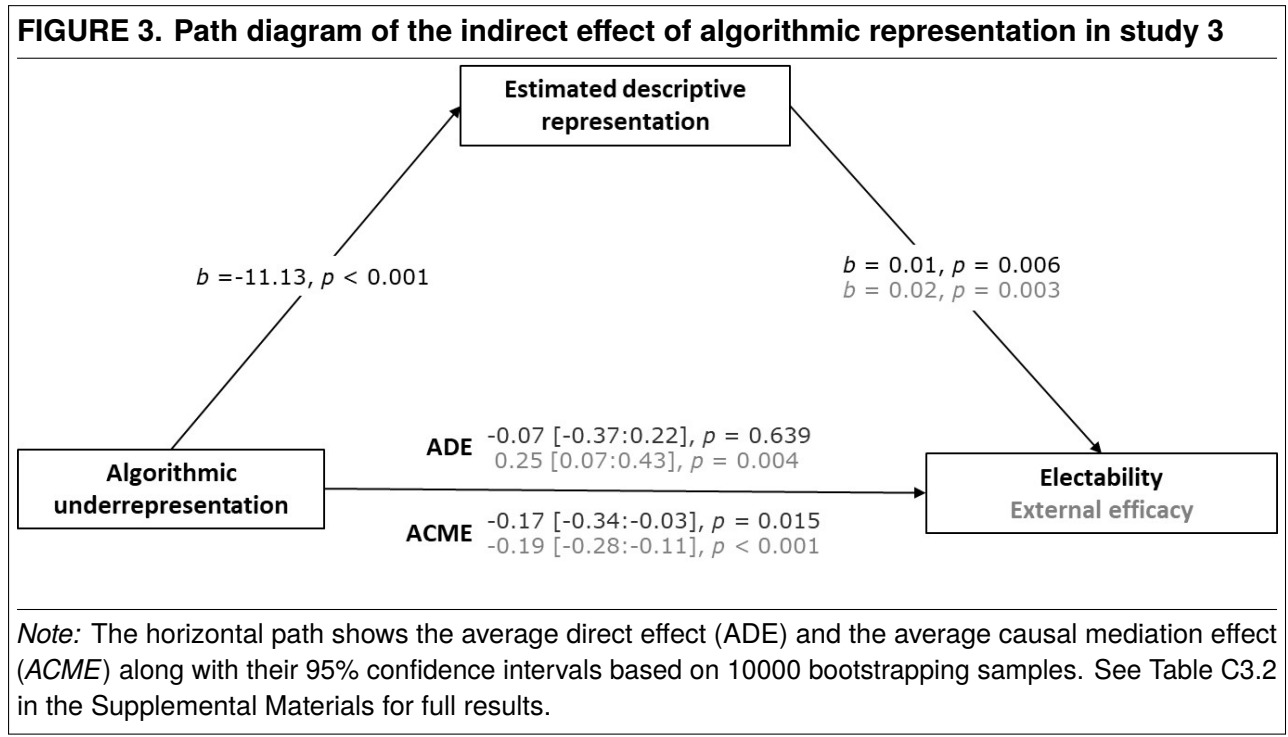

**FIGURE 3. Path diagram of the indirect effect of algorithmic representation in study 3**

*Note:* The horizontal path shows the average direct effect (ADE) and the average causal mediation effect (*ACME*) along with their 95% confidence intervals based on 10000 bootstrapping samples. See Table C3.2 in the Supplemental Materials for full results.

## Study 4: Consequences of algorithmic gender, race, and intersectional biases

The previous study focused on the specific case of women's representation in politics. The final study broadens the focus of the consequences of biased algorithmic representation to non-white and intersectional politicians (i.e. non-white women Philpot and Walton 2007; Stokes-Brown and Dolan 2010; Bateson 2020). Study 4 uses the same experimental "no bias" and "gender bias" conditions of the previous study but adds two more conditions: a "race bias" condition in which non-white politicians are underrepresented in Google outputs and an intersectional "double bias" condition in which both women and non-white politicians are underrepresented. The design of the stimuli for study 4 identical to that of study 3 except that some white persons were replaced with non-white persons. We again used Prolific to recruit a gender-balanced sample of 781 US participants.

Study 4 uses the same three measures but with a few additions. First, we drop the pre-treatment measure of *estimated descriptive representation* to avoid anchoring effects. Moreover, participants are now asked to estimate three shares—that of women, non-white (Black, Asian, or with Hispanic origins), and female non-white politicians. Second, we employ a more realistic 5-point question to measure perceived electability that is tailored to the study context (Bateson 2020; Corbett et al. 2022): "Do you think it will be harder or easier for a [woman/member with Black, Asian, or Hispanic origins] member to win reelections for the House of Representatives, compared to a [man/white member]?" In addition to this assessment of *general electability*, we now include a ranking exercise to capture perceptions of the *specific electability* of candidates. This exercise presented participants with a list of eight randomly ordered names that they were told represent candidates for primary elections of the party closest to them. The names were selected from a pretested list to convey a white or non-white male or female identity (two names per category; Crabtree et al. 2023). Participants then had to guess candidates' chances of winning the elections by dragging each name to a box with the labels high (=1), average (=0), or little (=-1) chances of winning the election. We then averaged the rankings of different name combinations to derive electability measures for (a) women, (b) non-white, and (c) non-white women candidates. Third, we use the same measures of external efficacy as in the previous study.

We first again assess the impact of exposure to algorithmic exposure on the perceived representation of minoritized groups (*H3*). We run separate linear regression models for the three estimation measures and insert the experimental conditions as dummy-coded predictors (along with control variables; see section C4 in the Supplemental Materials for the full models). Figure 4 illustrates the results of these models. It shows, on the one hand, that algorithmic representation with the balanced gender and race composition of depicted politicians results in slight overestimations of the descriptive representation of all minoritized groups. On the other hand, algorithmic underrepresentation of women and/or non-white politicians results in a consistent underestimation of their descriptive representation. As in the previous study, exposure to gender-biased Google output decreases the perception of women's presence in the U.S. House of Representatives ($b = -11.51$, $SE = 1.32$, $p < 0.001$). We find the same pattern for participants in the double bias condition where both women and non-white politicians are underrepresented ($b = -9.60$, $SE = 1.32$, $p < 0.001$). Conversely, the perceived descriptive representation of non-white politicians is negatively impacted by the algorithmic underrepresentation of non-white politicians (race bias: $b = -7.45$, $SE = 1.52$, $p < 0.001$; double bias: $b = -9.51$, $SE = 1.53$, $p < 0.001$). Finally, gender,

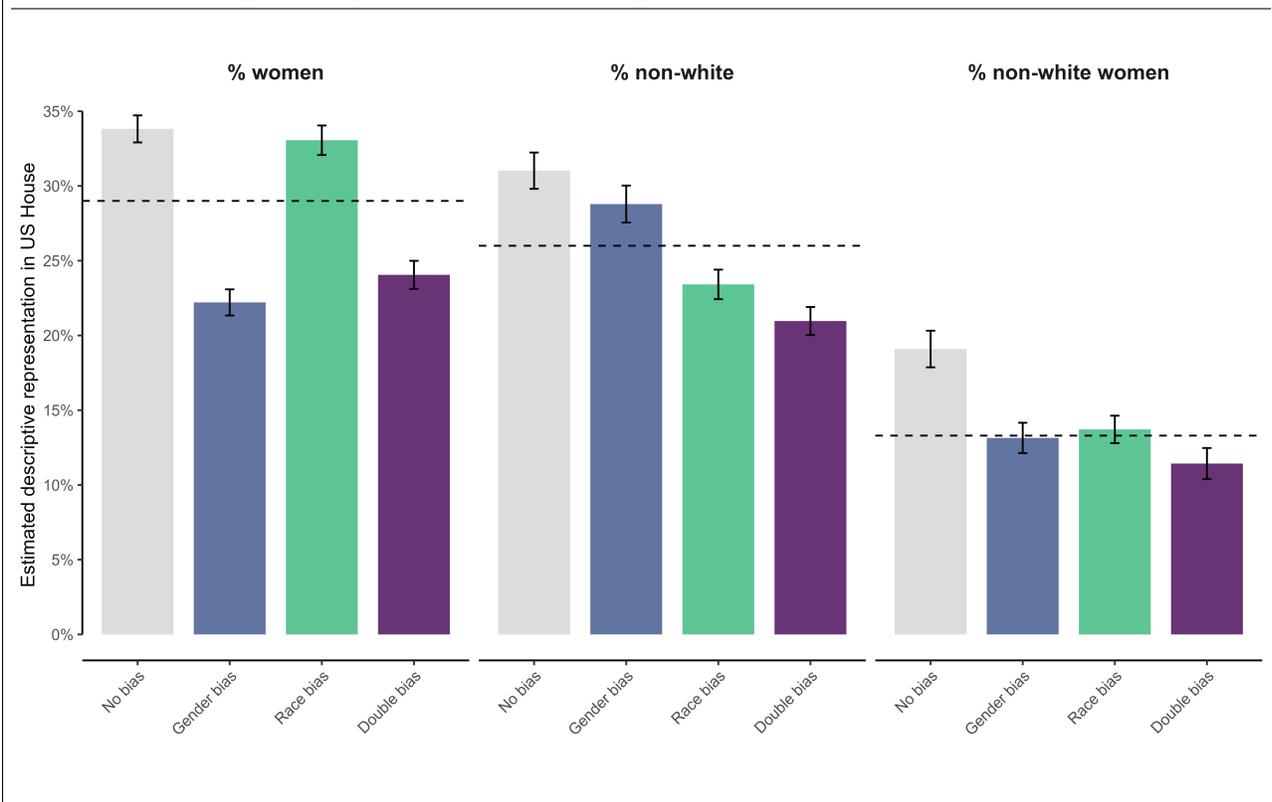

**FIGURE 4. Experimental effects of exposure to algorithmic representation on women's estimated descriptive representation in study 4**

*Note:* Experimental effects of exposure to algorithmic representation in study 4. Bars show the predicted post-treatment estimations of descriptive representation of women, non-white, and intersectional politicians across experimental conditions. The dashed lines denote the minoritized groups' actual descriptive representation. See Table C4.1 in the Supplemental Materials for full results.

race, and the intersectional bias in search engine representation all significantly lower the perceived presence of non-white women in the House of Representatives, compared to the balanced control condition. However, only the combined underrepresentation of women and non-white politicians results in a significant underestimation of intersectional politicians relative to their actual descriptive representation.

Finally, we look for evidence in our data for the causal strategic discrimination mechanism (*H4*). We conduct the same mediation analysis as in the study 3, except that we align the treatment and mediator with the minoritized group, resulting in a three-by-three grid (see Figure 5). For example, we test the treatment effect of the race bias (vs. no bias) condition on participants' assessments of non-white politicians' general and specific electability as well as participants' external efficacy. We find consistent evidence of the same causal pattern across all nine models (see blue rows in Figure 5):

The algorithmic underrepresentation of a group diminishes their perceived presence in politics which in turn translates into lower general and specific electability assessments as well as diminished beliefs in participants' external efficacy. As a last step, we explore whether the *ACME*s vary across different subgroups (see gray rows in Figure 5). Though the magnitude of effects varies between subgroups, their direction remains unchanged. Interestingly, the depressing effect of algorithmic (under-)representation on external efficacy beliefs extends to male (and to republican) respondents as well.

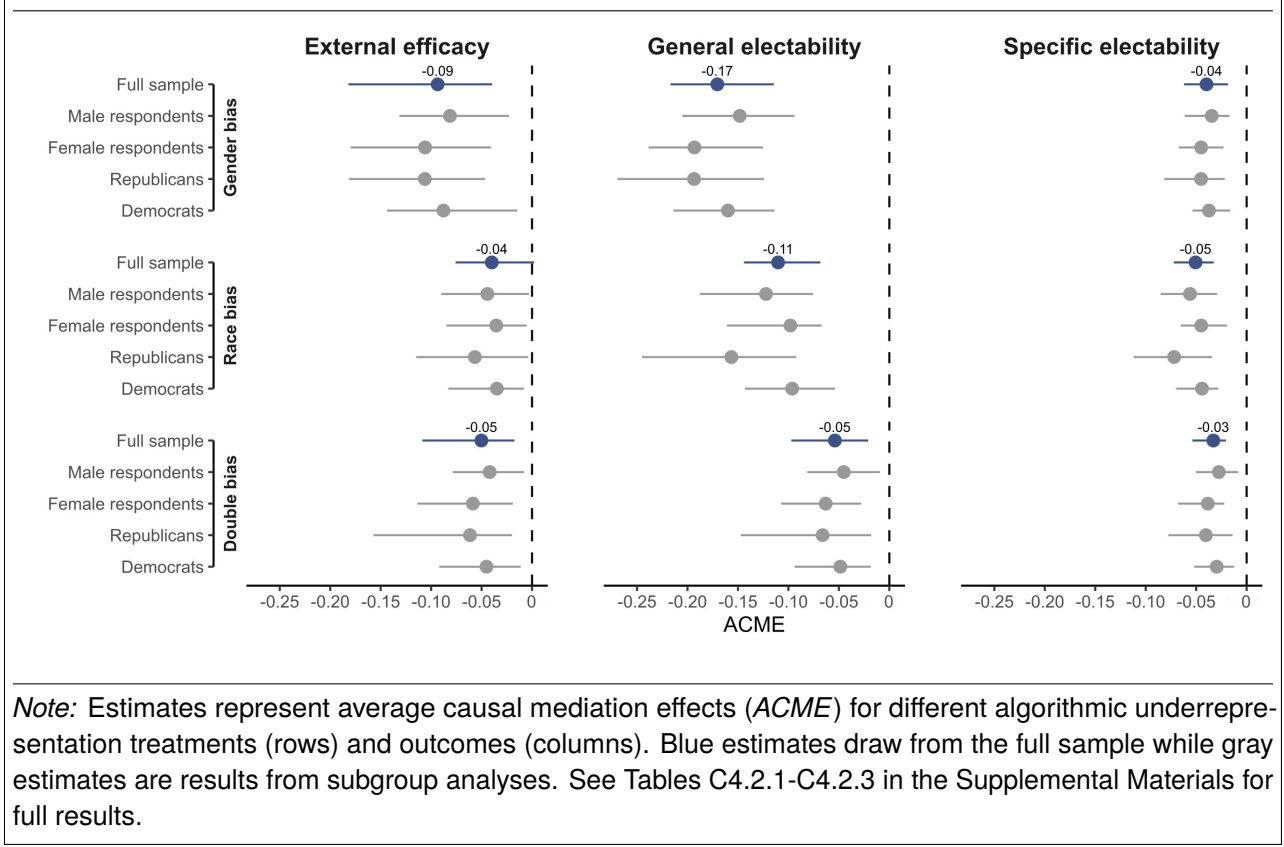

**FIGURE 5. Summary of indirect effects of algorithmic representation from a series of simple mediation analyses**

*Note:* Estimates represent average causal mediation effects (*ACME*) for different algorithmic underrepresentation treatments (rows) and outcomes (columns). Blue estimates draw from the full sample while gray estimates are results from subgroup analyses. See Tables C4.2.1-C4.2.3 in the Supplemental Materials for full results.

## GENERAL DISCUSSION & CONCLUSION

What is the role of search engines in upholding views of politics as a masculine and white domain? In this article, we proposed a framework of algorithmic representation of minoritized groups that theorizes this question as a chain of four intertwined types of bias, which we then empirically tested across four empirical studies. Overall, we find consistent support of all expectations. First, search engines algorithmically underrepresent women in their output in absolute terms. Second, the extent of women's algorithmic underrepresentation in Google image searches generally tracks with their collective descriptive representation in 84 legislative bodies across 56 countries. Third, exposure to gender, race, and intersectional bias in search engine outputs consistently distort perceptions of the political reality by decreasing the estimated presence of politicians belonging to these groups among individual search engine users. Finally, these misperceptions act as a casual conduit through which algorithmic biases not only diminish the perceived chances of winning an election for minority candidates but also result in voters feeling that their voices matter less.

Our findings have several implications for the study of representation in politics. First, our account of algorithmic representation introduces search engines as an important piece in the puzzle of persistent underrepresentation of minoritized groups in contemporary politics. The findings illustrate how widely used digital tools shape—and are shaped by—existing structural inequalities; they raise concerns about the future impacts of these tools, in particular considering the rapid adoption of AI-driven systems in different societal sectors. This combination of structural and algorithmic explanations of political bias goes with and beyond more established candidate- and voter-centered approaches (see also Dolan and Hansen 2018). On the supply-side, Juenke and Shah (2016) showed that voters can rarely "choose amongst a menu of racially or politically diverse candidates" (84). Our findings suggest that search engines may compound this dearth of minoritized candidates by rendering those who are running less visible through algorithmic filtering (see Thomsen and King 2020; O'Brien 2015). Moreover, googling politics and not finding any role models who share your gender or skin color might stymy nascent political ambition in potential future candidates (Bos et al. 2022; Fox and Lawless 2011; Kanthak and Woon 2015). On the demand-side, recent studies indicate that overt forms of voter bias are disappearing (Rohrbach et al. 2023; Schwarz and Coppock 2022; Teele et al. 2018). By conceptually broadening the notion of strategic discrimination (Bateson 2020; Green et al. 2022), we show that search engine outputs can act as the basis for voters' assessments of politicians' electability and thus present a more subtle and indirect form of voter bias. Empirically, our mediation analyses consistently support the notion of strategic discrimination as a causal driver of race- and gender-based prejudice in contemporary politics.

Second, we have theoretically and empirically outlined the circular logic of algorithmic bias in politics (see Savaget et al. 2019). Algorithms power AI models which are trained on data reflecting existing structural inequalities; in turn, algorithmic (mal)performance amplifies social and political realities through inherent biases in their filtering (Burrell and Fourcade 2021). All findings from our experimental conditions attest to this vicious cycle in which search engines underrepresent candidates from minoritized groups and thereby diminish their perceived electoral chances. However, this pattern reverses into a virtuous cycle in the conditions where we artificially removed bias from search engine output. Algorithmic overrepresentation of minoritized groups boosts voters' estimations of their political presence in government. This finding extends existing work on public perceptions of political udnerrepresentation (Stauffer 2021; Dolan and Hansen 2018; Dolan and Sanbonmatsu 2009) by positioning search engines as sources of public impression formation. In line with previous research (Stauffer 2021), the evidence linking perceived inclusion of minority groups to increased feelings of external efficacy is especially noteworthy. In times of fatiguing democratic support around the world (Diamond 2015; Gavras et al. 2022; Graham and Svolik 2020), algorithmic representation could play a vital role in fostering public support of political institutions.

Third, our empirical evidence contributes to ongoing public debates and cross-disciplinary research on algorithmic fairness (Kalluri 2020; Weinberg 2022; Wong 2020). Our findings are integral for increasing societal awareness about the discriminatory potential of AI-driven systems in the context of political participation (Savaget et al. 2019; Stier et al. 2022). By delineating the consequences of exposure to algorithmic representation, we showcase that the concept of "bias in, bias out" extends beyond a catchphrase; the functioning of proprietary AI-driven systems, such the ones used by Google, should be more actively incorporated into broader discussions on algorithmic governance and injustice (Birhane 2021). Accordingly, our findings provide an empirical basis for developing new regulation for preventing risks associated with the growing adoption of AI and is thus relevant for a broad range of stakeholders, including policymakers and industry, but also civil society and human right advocacy groups.

This study comes with several limitations. The algorithm audits center on a single case of Google image search. Insightful and widely used as this specific case may be, it remains unclear how our framework of algorithmic representation transfers to other AI-curated digital spaces. Future research should extend the focus to social media platforms like TikTok that heavily rely on AI to structure content resulting in the amplification of undesirable political messages, including hate and disinformation (Weimann and Masri 2023). Another limitation arises from the use of computer vision. Though manual coding suggests good reliability, this approach is limited to a binary classification and disregards the complexity of gender identity (Scheuerman et al. 2019). For ethical and baseline-related reasons regarding the use of AI, our auditing studies also ignore the question of representation of race. Critical analysis and in-depth qualitative work could help establish a benchmark for how search engines (mis)represent non-white politicians. Much of our analysis captures dynamics on the level of political institutions and groups of candidates. A promising approach would be extending our framework to the level of individual candidates to study algorithmic representation in the context of politicians with concrete political histories and, for instance, partisan identities (Pradel 2021). Finally, our experimental evidence documents algorithmic underrepresentation decreases perceived electability. However, more work is needed to delineate downstream electoral consequences of such diminished electability, ideally using observational data from actual elections or longitudinal designs. Whereas the political landscape is only slowly becoming more inclusive towards women and non-white politicians, the algorithmic landscape has been evolving rapidly in all directions. This article captured how these two landscapes are interlocked to sustain white and masculine views of politics at present; tracking how these dynamics unfold in the future remains a major concern for the legitimacy of democratic structures and processes.

Algorithmic gender and race biasWomen's Underrepresentation in Elections". *American Journal of Political Science* 68 (1): 289–303.

Atkeson, Lonna Rae and Nancy Carrillo. 2007. "More is Better: The Influence of Collective Female Descriptive Representation on External Efficacy". *Politics & Gender* 3 (1): 79–101.

Bandy, Jack. 2021, April. "Problematic Machine Behavior: A Systematic Literature Review of Algorithm Audits". *Proceedings of the ACM on Human-Computer Interaction* 5 (1): 1–34.

Bast, Jennifer. 2024, January. "Managing the Image. The Visual Communication Strategy of European Right-Wing Populist Politicians on Instagram". *Journal of Political Marketing* 23 (1): 1–25.

Bateson, Regina. 2020. "Strategic discrimination". *Perspectives on Politics* 18 (4): 1068–1087.

Birhane, Abeba. 2021, February. "Algorithmic injustice: a relational ethics approach". *Patterns* 2 (2): 1–9.

Bos, Angela L, Jill S Greenlee, Mirya R Holman, Zoe M Oxley, and Celeste J Lay. 2022. "This One's for the Boys: How Gendered Political Socialization Limits Girls' Political Ambition and Interest". *American Political Science Review* 116 (2): 484–501.

Bratton, Kathleen A and Leonard P Ray. 2002. "Descriptive representation, policy outcomes, and municipal day-care coverage in Norway". *American Journal of Political Science* 46 (2): 428–437. Type: Journal Article.

Burrell, Jenna and Marion Fourcade. 2021, July. "The Society of Algorithms". *Annual Review of Sociology* 47 (1): 213–237.

Carpinella, Colleen and Nichole M. Bauer. 2021, March. "A visual analysis of gender stereotypes in campaign advertising". *Politics, Groups, and Identities* 9 (2): 369–386.

Cavazos, Jacqueline G, Jonathon P. Phillips, Carlos D. Castillo, and Alice O. O'Toole. 2021. "Accuracy Comparison Across Face Recognition Algorithms: Where Are We on Measuring Race Bias?". *IEEE Transactions on Biometrics, Behavior, and Identity Science* 3 (1): 101–111.

Corbett, Christianne, Jan G Voelkel, Marianne Cooper, and Robb Willer. 2022. "Pragmatic Bias Impedes Women's Access to Political Leadership". *Proceedings of the National Academy of Sciences* 119 (6): 1–11.

Crabtree, Charles, Jae Yeon Kim, S. Michael Gaddis, John B. Holbein, Cameron Guage, and William W. Marx. 2023, March. "Validated names for experimental studies on race and ethnicity". *Scientific Data* 10 (1): 130.
23

Algorithmic gender and race bias